\documentclass[aps]{article}
\usepackage{geometry}              
\usepackage{graphicx}
\usepackage{subfig}
\usepackage{amssymb}
\usepackage{epstopdf}
\usepackage{amsmath}
\usepackage{fixmath}
\usepackage{MnSymbol}
\usepackage{amsfonts}
\usepackage{bm}
\DeclareMathAlphabet{\mathpzc}{OT1}{pzc}{m}{it}

\title{Objective Nontensor Rheology: Unique Flow Decompositions from Correlated Microscopic Motions}
\author{Clifford Chafin\\
\small{Department of Physics, North Carolina State University, Raleigh, NC 27695} \thanks{cechafin 'at' ncsu.edu}}

\begin{document}
\maketitle
\begin{abstract}
The use of continuum mechanics and invariants built from the deviator as an adequate foundation for rheology has been recently disputed by this author.  Here we give a specific example of the kind of parcel deformations that are uniquely decomposed by way of microscopic motions into a maximal rotation, a pure shear and an extension.  The construction of these equations depends on only one free material parameter but they have no nice form in terms of the operations of vector and tensor calculus which may be why they were overlooked.  Although the first order flow is often sufficient to give the rheological information, finite sized parcel deformations can give confusion because of boundary effects, the relevance of which are highly dependent on the global geometry of the experiment.  
\end{abstract}

Rheology is the study of hydrodynamics when things get interesting.  Simple fluids obey the Navier-Stokes equations with a shear and (usually zero) bulk viscosity.  This reduces the hydrodynamics to a function of density, velocity, these two viscous parameters and, sometimes some thermodynamic variables \cite{Batchelor}.  For gases the N-S equations follow directly from the Maxwell-Boltzmann distribution.  This case is almost too gratifying and, for those of us who like trouble, it is a great comfort that higher order approximations diverge \cite{Dorfman}.  The mathematician in us loves the idea that more complex fluids can be fit into a simple tensor expansion that reflects some underlying symmetries and conservation laws of the fluid.  In the case of gases, this is wonderfully spoilt by small scale fluctuations and the problems of time delay from the finite collision time reminiscent of the radiation reaction problem, the heat equations and the effects of finite time delay there. There are attempts to improve results by using resummation methods but a fundamental understanding of higher order hydrodynamics of gases is still lacking.  Liquids are essentially different in that the stresses are not mediated by diffusive momentum transfer over several times the interparticle spacing but by bond strain as in solids but of a transient nature.  

In a recent work this author showed that the highly correlated microscopic motions of liquids allow us to average out the diffusive motion of particles and that these then define a unique local notion of shear, extension and rotation of each local parcel \cite{Chafin-rh}.  This is in contrast with a famous result of Helmholtz that a shear is equivalent to an extension and a rotation \cite{Bird, Truesdell}.  From this we concluded that we should not simply speak of ``shear viscosity'' as covering both linear shear and extension motions but have separate distinctions for each and have introduced a change in nomenclature for each (so that ``shear'' will refer only to the unique parallel plate type flow we extract from our decomposition).  The previous result was done with an appeal to how forced deformation and a finite granularity scale must drive mixing of solutions with varying concentration gradients.  Here I present a simple example that illustrates this.  

\begin{figure}
\begin{center}
\includegraphics[trim = 0mm 100mm 10mm 30mm, clip,width=3in]{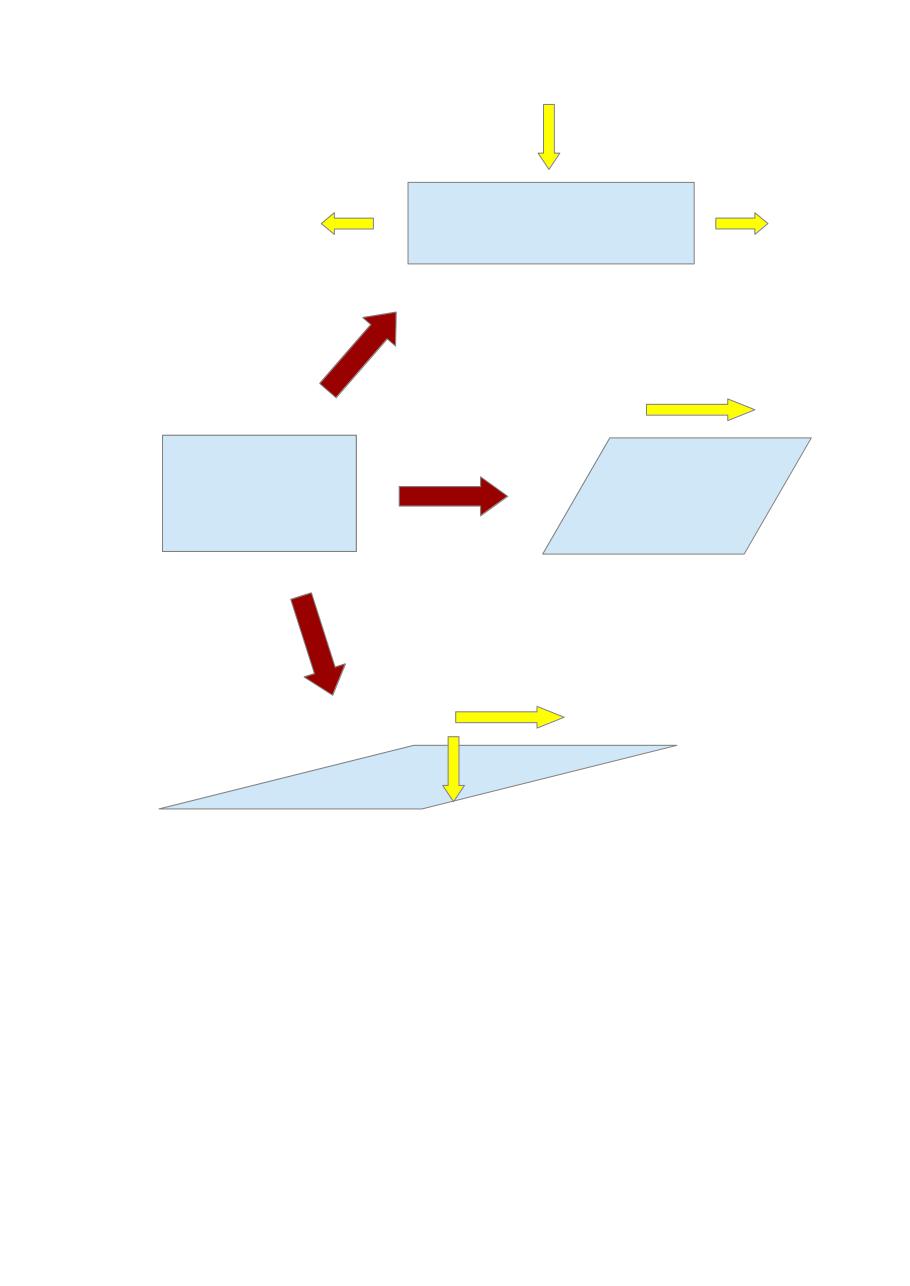}
\caption{A finite sized fluid parcel under volume conserving compression, linear pure shear and shear with extension.}
\label{shear}
\end{center}
\end{figure}

Consider a finite sized rectangular parcel of fluid as in fig.\!~\ref{shear}.  Since each parcel is made of thermally diffusing finite sized units, we can consider the upper deformation as a volume conserving extension flow.  The middle deformation is
akin to a sequence of billiard balls sliding past each other.  The number of balls stays the same  so we may consider the volume to be fixed but the surface bounding the volume to have grown.  (The coordinated number of ``bonds'' at the surface has decreased.)  What is confusing here is that, in the continuum limit, this sequence of discrete steps is obscured and seems like a smooth diagonal slope which greatly underestimates the area!  This deformation is apparently highly reversible.  Such a parcel looks like what we expect in a shear plate rheometer sample or in Couette flow.  The periodicity of such a situation eliminates the surface area problem we obtain by using a discrete rectangular sample.  

In the bottom case, we have a volume preserving deformation where the the whole parcel flattens down to a long thin layer that is a combination of both shear and extension.  In the limit of a single layer thick result the particles did not just have to slide reversibly past each other but must force interstitial step choices that cannot be undone predictably.  If we have a concentration gradient vertically in the sample we see that this one leads to irreversible mixing and no reversed deformation will undo it.  We conclude that this motion is a combination of a lateral shear and a vertical (inwards) extension.  Interestingly, the bounding surface area has no such underestimate as above.  
Mixing alone is not a determination of viscosity but, since the deformation drives orientation and stretch of long particles as a function of their relaxation times, this alters both mixing and  the kind of bond stress possible mediated in the fluid.  

The ultimate goal of physics is to see how microscopic dynamics determines macroscopic physics.  Since the usual continuum mechanical approach can put invariants together in terms of the deviator $D_{ij}$ and these cannot distinguish such effects, it seems evident that no analytic expressions or expansions in terms of invariants of these\footnote{``Invariant'' in the sense that no finite rotation, translation or boost alters the form of the equations. Finite velocity rotations are not classical invariants but, for the small parcel sizes in fluids, this is locally valid as well. } can generally give an adequate rheology.  The microscopic motions are certainly not as simple as billiard balls.  In fact, we require that diffusive local motion be faster than that induced by deformations to be a fluid at all.  Since a rotational velocity of a parcel gives negligible centrifugal force, it can induce no forces on the system.  Internal forces come from frustrating/biasing the diffusive motion.  For nontrivial motion, by tracking the local motions about a particle one finds that there is generally a maximal rotational component about some axis.  The rest can be locally described by some shear defined by the axis giving a maximal curl component.  After removing the rotation and shear the velocity field is irrotational and uniquely defines a local linear extension about each poin 
\begin{align}
v=v_{0}+v_{r}+v_{s}+v_{e}
\end{align}
where $v_{0}$ is the velocity at the point and the rest are linear order approximations in its neighborhood.  

In practice this separation, while physically meaningful and necessary, is not so lovely and may be why it has been overlooked until now.  Mathematical aesthetics have far intruded into continuum mechanics; so much so that some may have forgotten that they are never a sufficient answer on their own, regardless of how convenient a shortcut they introduce.  
Looking at shear flow $v_{x}=x$ versus rotation $v_{\theta}=c r$, which both have a spatially uniform curl, and Helmholtz decomposition theorem
\begin{align}
v&=\nabla\phi+\nabla\times A\\
\phi&=\frac{1}{4\pi}\int_{V}\frac{\nabla'\cdot v(r')}{|r-r'|}dV'-\frac{1}{4\pi}\oint_{S}n'\cdot\frac{v(r')}{|r-r'|}dS'\\
A&=\frac{1}{4\pi}\int_{V}\frac{\nabla'v(r')}{|r-r'|}dV'-\frac{1}{4\pi}\oint_{S}n'\times\frac{v(r')}{|r-r'|}dS'
\end{align}
we see that the boundary contributions are essential to distinguish these so that the curl and extension (the irrotational contribution) is not enough local information to address this problem.  It is disappointing and a bit counterintuitive, given the predominant role vector calculus plays in our education of fluids, that the div, grad, curl and the family of associated identities is inadequate to describe something essential about local flow for rheology.  
Since the local curl is unambiguous, we expect the shear and rotation to be about the same axis.  We will see momentarily there is a one parameter closed interval of choices for the ratio of how much of each is manifested microscopically for a given flow.  This will be represented by a material parameter associated with the flow.  

A procedure to create this decomposition should give a maximal rotational component and minimize the shear flow however this turns out not give a completely unique characterization.  Considering a mixture of shear and rotation in 2D it is evident that the vector field projection along an $\epsilon$-sized circle about a point will have a sinusoidal single period wobble when the shear is nonzero.  This is removed by a unique oscillatory counterflow that leave a uniform vector field corresponding to rotation.  If we project the vector field on $\hat{\theta}$ as $s=v\cdot\hat{\theta}$ we can see that this function is unchanged if we use a 180$^{\circ}$ square step filter $F=\Theta(\pi-\theta)\cdot\Theta(\theta)$ as $s(\theta)=\int\ d\theta' s(\theta')F(\theta'-\theta-\pi/2)$.  However, any projection from an extension of the form $\mathpzc{s}=\hat{\theta}\cdot(\nabla\phi-v_{0})$ is annihilated.  This can be seen from the fact that these are equivalent to multipolar distributions that are quadrupolar or higher.  


Using this observation we can compute the decomposition as follows.  Use $c=\nabla\times v$ to define a perpendicular plane $P_{\perp}$ with variables $r,\psi$.  Project the vector field on the plane by $v^{\perp}=v-v\cdot\hat{c}$.  Use our filter $s'=\mathcal{F}(s)$ on the function $s=\frac{1}{\epsilon}\hat{\theta}\cdot v^{\perp}(\epsilon,\theta)$ to remove any extension component of the flow.  This function $s'(\theta)$ has max and min $M$ and $m$.  These occur at angles $\theta_{M}$ and $\theta_{m}=\theta_{M}+\pi/2$ (corresponding to coordinate axes $x'$ and $y'$ in $P_{\perp}$).  We can choose any decomposition with rotation rate $\rho\in [m,M]$ and shear rates $h_{1}=\rho-M$ and $h_{2}=\rho-m$.  Define the material parameter $\alpha=\frac{\rho}{M-m}$ as a measure of how the medium divides shear and rotation in a given local linear flow.  This gives the corresponding velocities
\begin{align}
v_{r}&=\rho r \hat{\theta} \\
v_{s}&=h_{1}  y' \hat{x}'  +h_{2} x' \hat{y}'
\end{align}
The extension flow is the remainder of the flow $v_{e}=v-v_{0}-v_{r}-v_{e}=\nabla\psi$.  Since $v_{e}$ is linear in the local variables $\psi$ is quadratic and the extension axes are the directions where $v_{e}$ is parallel to $\hat{R}$, the radial coordinate in 3-space about the point.  

%
%


The advantages of such a decomposition is that it gives variables adequate to describe the deformation driven mixing, which must primarily\footnote{Shear can alter lateral diffusion rates but this is a much smaller effect.} depend on $v_{e}$, and lets us explicitly remove $v_{r}$ from any discussion of viscosity.  This meets the objective criterion in the most direct possible way, with no ambiguities introduced by abstract, and probably physically irrelevant, constructions, like the polar decomposition.  Interestingly, even nonlinear contributions can be included directly from $u=v-v_{r}-v_{0}$ to any order in a manifestly objective fashion.  Specifically, any induced stress tensor of the form $T=P\cdot I +f(u)$ which is symmetric and gives a damping of energy is physical, objective and obeys conservation laws.  The case of objective and conserving retarded effects is not so clear.  The best solution may be to introduce hidden internal variables like bond strain, coordination or constituent particle stretch so that no history dependence is needed.  The connection between the microscopic motions and regional flow is then reduced to finding the class of typical local particle configurations for a given local linear flow profile and slowly changing internal parameters.  The presence of locally well defined axes and planes for rotation, shear and extension means that we pull the parameters for response directly from the local ensembles of sheared particles.  It is very hard to see how one could make this association to the case of models built on invariants in any way other than curve fitting.  

One needs to wonder how hard one would have to drive a fluid to get higher order flow curvature effects and true nonlinearity.  The extent of polymers is almost always microscopic on a scale where fluid is changing appreciably.  This suggests that when internal variables are included that the local linear flow information is completely adequate for any  practical rheology.  Hiding these internal states by trying to represent them as higher order or retarded info is possible but seems less directly connected to the realistic microscopic state of the fluid.  Although the statement will undoubtably shock some, the author suggests that, except for microfluidics and extreme turbulence, rheology never needs any flow information beyond $W_{ij}=\partial_{i}v_{j}$ for purposes of viscosity and mixing and all apparently nonlinearity, history and higher order effects should be relegated to dynamics of new local variables that depend only on $W$ and each other.


\end{document}